\documentstyle[12pt]{article}
\newcommand{\nsect}{\setcounter{equation}{0}
\def\theequation{\thesection.\arabic{equation}}\section}

\begin{document}
\newcommand{\newc}{\newcommand}
\newc{\R}{$R$}
\newc{\ru}{$U(1)_R$\ }
\newc{\fun}[4]{\kz{1}^{#1}\kz{2}^{#2}\kz{3}^{#3}\kz{4}^{#4}}
\newc{\funth}[3]{\kz{1}^{#1}\kz{2}^{#2}\kz{4}^{#3}}
\newc{\funtw}[3]{\kz{1}^{#1}\kz{3}^{#2}\kz{4}^{#3}}
\newc{\funfo}[3]{\kz{1}^{#1}\kz{2}^{#2}\kz{3}^{#3}}
\newc{\funtwth}[2]{\kz{1}^{#1}\kz{3}^{#2}}
\newc{\del}{\partial}
\newc{\vev}[1]{<\!{#1}\!>}
\newc{\beq}{\begin{equation}}
\newc{\eeq}{\end{equation}}
\newc{\barr}{\begin{eqnarray}}
\newc{\earr}{\end{eqnarray}}
\newc{\ra}{\rightarrow}
\newc{\lam}{\lambda}
\newc{\eps}{\epsilon}
\newc{\half}{\frac{1}{2}}
\newc{\third}{\frac{1}{3}}
\newc{\fourth}{\frac{1}{4}}
\newc{\eighth}{\frac{1}{8}}
\newc{\gev}{\,GeV}
\newc{\lra}{\leftrightarrow}
\newc{\Dslash}{\not\!\! D}
\newc{\sg}{{\cal G}}
\newc{\eq}[1]{(\ref{eq:#1})}
\newc{\eqs}[2]{(\ref{eq:#1},\ref{eq:#2})}
\newc{\etal}{{\it et al.}\ }
\newc{\Hbar}{{\bar H}}
\newc{\hhbar}{{\overline h}}
\newc{\Ubar}{{\bar U}}
\newc{\Dbar}{{\bar D}}
\newc{\Ebar}{{\bar E}}
\newc{\eg}{{\it e.g.}\ }
\newc{\ie}{{\it i.e.}\ }
\newc{\nonum}{\nonumber}
\newc{\kap}{\kappa}
\newc{\kapi}{\frac{1}{\kap}}
\newc{\kz}[1]{(\kap z_{#1})}
\newc{\lab}[1]{\label{eq:#1}}
\newc{\oc}{{\cal O}}
\newc{\vecl}{\vec{l}}
\newc{\lle}[3]{L_{#1}L_{#2}\Ebar_{#3}}
\newc{\lqd}[3]{L_{#1}Q_{#2}\Dbar_{#3}}
\newc{\udd}[3]{\Ubar_{#1}\Dbar_{#2}\Dbar_{#3}}

\title{Anomaly-Free Gauged \R-Symmetry in Local Supersymmetry}
\author{A. H. Chamseddine and Herbi Dreiner}
\date{{\small Theoretische Physik, ETH-H\"onggerberg, CH-8093 Z\"urich,
Switzerland}}
\maketitle

\vspace{-6.5cm}
\hfill\parbox{8cm}{\raggedleft \today \\ ETH-TH/95-04}
\vspace{6.5cm}

\begin{abstract}
\noindent We discuss local \R-symmetry as a potentially powerful new model
building tool. We first review and clarify that a $U(1)$ \R-symmetry can only
be gauged in local and not in global supersymmetry. We determine the
anomaly-cancellation conditions for the gauged \R-symmetry. For the standard
superpotential these equations have {\it no} solution, independently of how
many Standard Model singlets are added to the model. There is also no solution
when we increase the number of families and the number of pairs of Higgs
doublets. When the Green-Schwarz mechanism is employed to cancel the
anomalies, solutions only exist for a large number of singlets. We find many
anomaly-free family-independent models with an extra $SU(3)_c$ octet chiral
superfield. We consider in detail the
conditions for an anomaly-free {\it family dependent} $U(1)_R$ and find
solutions with one, two, three and four extra singlets. Only with three and
four extra singlets do we naturally obtain sfermion masses of order the
weak-scale. For these solutions we consider the spontaneous breaking of
supersymmetry and the $R$-symmetry in the context of local supersymmetry. In
general the $U(1)_R$ gauge group is broken at or close to the Planck scale. We
consider the effects of the \R-symmetry on baryon- and lepton-number violation
in supersymmetry. There is no logical connection between a conserved
\R-symmetry and conserved \R-parity. For conserved \R-symmetry we have models
for all possibilities of conserved or broken \R-parity. Most models predict
dominant effects which could be observed at HERA.
\end{abstract}

\nsect{Introduction}
Supersymmetry combines fields of different spin into supermultiplets. It
includes the special possibility of a symmetry which distinguishes between the
fermionic and the bosonic component of a $N=1$ super\-symmetric superfield.
Such sym\-metries are called \R-sym\-metries and they are particular to
supersymmetry. As such, they deserve special attention when considering the
implications of supersymmetry. \R-parity can be thought of as a discrete
\R-symmetry and has been widely discussed in the context of the minimal
supersymmetric standard model and its extensions. There is also a considerable
amount of literature on global \R-symmetries and their phenomenological
implications. It is the purpose of this paper to reconsider local
\R-symmetries in the context of local supersymmetry and to make first steps
towards a realistic model. It is similar in spirit to \cite{us} where we
considered a non-R $U(1)'$.

\R-symmetries were first introduced in {\it global} supersymmetry by Salam and
Strathdee \cite{rinvariance1} and by Fayet \cite{rinvariance2} in order to
enforce global lepton- or baryon-number. In the following years, the discrete
symmetry \R-parity \cite{rp} has been imposed to prohibit all dimension four
lepton- and baryon-number violating interactions which arise in the
supersymmetric extension of the Standard Model. Global \R-invariance has been
proposed as a solution to the strong CP-problem \cite{buwy}, the mu problem
\cite{kimnil,dine,hara}, and the problem of the neutron electric dipole moment
\cite{buwy,affleck}. Global \R-invariance prohibits tree-level gaugino masses.
This leads to the interesting possibility that the gaugino masses are
generated radiatively or through a dynamical mechanism and thus predicted
\cite{barb,hara,fama}. If the global \R-symmetry remains unbroken to low
energies \cite{hara} then only the electroweak  gaugino masses can be
generated after $SU(2)_L\times U(1)_Y$ breaking. (A bi-scalar mu-term must be
generated or inserted by hand in the soft-susy breaking sector.) The radiative
gluino mass is very light \cite{barb,fama} and  excluded \cite{lightgluino}. A
heavy gluino can be obtained by adding an $SU(3)_c$ octet chiral superfield
\cite{hara}. One then loses any prediction for the gluino mass. This is not
very natural but it is consistent with experiment. However, the potential of
the scalar component of the octet is necessarily unrestricted and typically
breaks $SU(3)_c$. If the global \R-invariance is spontaneously broken
\cite{fama} one has an unwanted light pseudo Goldstone boson. The gaugino
masses can still be generated radiatively and the gluino is light \cite{fama}.
One can add explicit \R-breaking terms which give mass to the axion. However,
if these terms are large this renders the \R-symmetry meaningless. Recently
global \R-symmetries have been seen to arise in so-called generic models of
global supersymmetry breaking \cite{nelsei}. The problems of the axion from
\R-breaking are resolved when embedded into local supersymmetry through
explicit breaking terms \cite{bagpop}. Thus models with global \R-symmetry
suffer from an axion or a light gluino problem.

Beyond the immediate phenomenological problem of constructing a model with
global \R-symmetry there is a more fundamental problem. Supersymmetry
breaking is necessarily embedded in local supersymmetry. Local
supersymmetry automatically includes gravity and global symmetries are most
likely broken by quantum gravity effects \cite{gravglob}. Thus at low
energies we do not expect global symmetries such as baryon- or
lepton-number to be fundamental symmetries of nature but only symmetries of
the low-energy effective Lagrangian. At high-energy we expect all relevant
symmetries to be gauge symmetries. We shall thus investigate whether an
\R-symmetry can be gauged. We  provide a new local symmetry as a
model-building tool.

\medskip

Our paper is structured as follows. In Section 2 we show that an \R-symmetry
can only be gauged in local supersymmetry. In Section 3 we then consider
the conditions for an anomaly-free gauged \R-symmetry and find several
solutions. In Section 4 we discuss the spontaneous breaking of the
\R-symmetry and of supersymmetry. We find the important result that the
\R-symmetry is always broken at or near the Planck scale. In Section 5 we
consider the implications for \R-parity violation. In Section 6 we offer
our conclusions and an outlook.

\nsect{R-invariant Supersymmetric Theories}
Below we first discuss \R-symmetries in global supersymmetry  and in the
following subsection extend the discussion to local supersymmetric
theories, where \R-symmetries have not been as widely discussed.

\subsection{Global Supersymmetry}
For globally supersymmetric theories the global \R-transformations
are defined as a transformation on the superfields \cite{rinvariance1}
\beq
\begin{array}{ccc}
V_k(x,\theta,{\bar\theta})& \ra&
V_k(x,\theta e^{-i\alpha},{\bar\theta}e^{i\alpha}), \\
S_i(x,\theta,{\bar\theta}) &\ra& e^{in_i\alpha}
S_i(x,\theta e^{-i\alpha},{\bar\theta}e^{i\alpha}),
\end{array}
\lab{grtrans}
\eeq
where $V_k$ is a gauge vector multiplet with components
$V^\mu_k,\,\lam_k$,
and $D_k$ and $S_i$ are left-handed chiral superfields with components
$z_i,\chi_i$, and $F_i$. Thus the Grassman coordinates
$\theta,{\bar\theta}$ have non-trivial \R-charge
\beq
\theta\ra e^{-i\alpha}\theta,\quad{\bar\theta}\ra e^{i\alpha}{\bar\theta},
\quad \int \!d\theta\ra e^{i\alpha}\int \!d\theta,\quad
\int \!d{\bar\theta}\ra e^{-i\alpha}\int \!d{\bar\theta}.
\lab{grassman}
\eeq
The latter two transformations hold since for Grassman variables
integration is like differentiation. The \R-transformations act on the
components of the superfields as
\beq
\begin{array}{ccc}
V_k^\mu&\ra& V_k^\mu, \\
(\lam_k)_L&\ra&\exp(-i\alpha)(\lam_k)_L, \\
(\lam_k)_R&\ra&\exp(i\alpha)(\lam_k)_R, \\
D_k&\ra& D_k,
\end{array} \qquad
{\rm and}\qquad
\begin{array}{ccc}
z_i&\ra& \exp(in_i\alpha)z_i, \\
\chi_i&\ra&\exp\{i\gamma_5(n_i-1)\alpha\}\chi_i,\\
F_i&\ra&\exp\{i(n_i-2)\alpha\}F_i.
\end{array}
\lab{comptrans}
\eeq
We see that all gauginos transform non-trivially and with the {\it same}
charge. The scalar fermions transform differently from their fermionic
superpartners as we expect for an \R-symmetry. However, different chiral
supermultiplets will in general have different \R-charge, and
\R-transformations are more general than lepton- or baryon-number. The action
for the superpotential
\beq
\int d^2\theta\,g(S_i),
\eeq
is invariant provided that the superpotential transforms as
\beq
g(S_i)\ra e^{-2i\alpha}g(S_i),
\lab{rsuper}
\eeq
under the transformation \eq{grtrans}. Here we have made use of \eq{grassman}.
We see that the superpotential transforms non-trivially. This is one essential
fact of \R-symmetries. The kinetic terms of the vector and scalar multiplets
are of the form
\beq
\int d^2\theta d^2{\bar\theta}
\left[({\overline S} e^{2{\tilde g}V})^iS_i \right]
+ \int d^2\theta W^\beta W_\beta + \int d^2{\bar\theta}
{\overline W}_{\dot{\beta}}{\overline W}^{\dot{\beta}},
\lab{globalaction}
\eeq
where
\beq
W_\beta={\overline D}^2(e^{-V}D_\beta e^{V}),\quad
{\overline W}_{\dot{\beta}}=D^2(e^{-V}{\overline D}_{\dot{\beta}}e^V),
\eeq
are automatically invariant under \eq{grtrans}, \ie\
\beq
W_\beta\ra e^{-i\alpha}
W_\beta, \quad {\overline W}_{\dot{\beta}}\ra e^{i\alpha}{\overline W}_
{{\dot\beta}}.
\eeq
In this paper, we focus on the possibility of locally \R-symmetric
theories. However, as we now discuss it is {\it not} possible
in globally supersymmetric theories to promote the global \R-invariance
to a local one. An easy way to see this is to notice that when the
\R-parameter $\alpha$ becomes $x$-dependent then the transformations
\eq{grassman} change to
\beq
\theta\ra\theta e^{-i\alpha(x)},\quad {\bar\theta}\ra{\bar\theta}
e^{i\alpha(x)},
\eeq
which is a special form of a {\it local} superspace transformation.

In more detail, one can also see this from Eq.\eq{comptrans} which implies
that all gauginos have \R-charge, including the \R-gauginos. If the
\R-symmetry is to become a local symmetry then the \R\  gauge vector boson
$V_\mu^R$ will have to couple to the \R-gauginos $\lam^R$ in the form
\footnote{The lower index $(L,R)$ on the gaugino $\lam$ is the chirality
and the upper index \R\  indicates the gauge group.}
\beq
{\cal L}\sim
{\overline \lam^R_L}(\del_\mu-ig_R V_\mu^R)\gamma^\mu\lam_L^R
+{\overline \lam^R_R}(\del_\mu+ig_R V_\mu^R)\gamma^\mu\lam_R^R,
\lab{rgaugino}
\eeq
since the opposite chirality gauginos have non-trivial and opposite
\R-charge \eq{comptrans}. The above equation implies the coupling
\beq
{\cal L}\sim
g_R{\overline\lam^R}\gamma^\mu\gamma_5\lam^R V_\mu^R,
\lab{rterm}
\eeq
in the Lagrangian which is an axial interaction and is not present in
the action \eq{globalaction}. In order to construct a supersymmetric
Lagrangian containing \eq{rterm} we must consider its
supersymmetric transformation. It contains the
term\footnote{We make use of the identity $\gamma_\alpha\gamma_\beta
\gamma_\lam=g_{\alpha\beta}\gamma_\lam+g_{\beta\lam}\gamma_\alpha
-g_{\alpha\lam}\gamma_\beta+i\eps_{\mu\alpha\beta\lam}\gamma^\mu\gamma^5$.}
\beq
g_R\eps^{\mu\nu\rho\sigma}{\overline\eps}\gamma_\mu\lam^R
F_{\nu\rho}^RV_\sigma^R=\eps^{\mu\nu\rho\sigma}\delta V_\mu^RV_\nu^R
F_{\rho\sigma}^R,
\eeq
since the supersymmetric variation of the gaugino term $\delta\lam^R$
contains $\gamma^{\mu\nu}\eps F_{\mu\nu}^R$. $\eps$ is the infinitesimal
parameter of the supersymmetry transformation. The above term can not be
cancelled without departing from the setting of global supersymmetry.

{}From Eq.\eq{comptrans} it is clear that the \R-symmetry generator \R\
does not commute with the supersymmetry generator $Q$. In the literature
this is quoted as an argument that an \R-symmetry can not be gauged.
Explicitly we have \cite{west}
\beq
\left[Q_\alpha,R \right] = i(\gamma_5)_\alpha^\beta Q_\beta.
\eeq
Thus \R-symmetry is an extension of supersymmetry with the chiral
generator
and the extension is a graded Lie Algebra. If the \R\  generator of a
globally \R-supersymmetric theory is promoted to a local symmetry then
the above equation can only hold if the transformation parameters of the
supersymmetry algebra are $x$-dependent, \ie $Q_\alpha$ is the generator
of a local transformation.

Thus  \R-symmetries are intimately connected with supersymmetry:
a {\it locally} \R-invariant theory can only be constructed in a
{\it locally} supersymmetric framework; in {\it global} supersymmetry
only {\it global} \R-symmetries can be constructed.

\subsection{Local Supersymmetry}
In local supersymmetry the field content is extended to include a spin 2
graviton and a spin $\frac{3}{2}$ gravitino. From Eq.\eq{comptrans} we
generalize the \R-symmetry to the graviton multiplet as
\beq
\begin{array}{ccc}
e^m_\mu&\ra& e^m_\mu, \\
\psi_\mu&\ra&\exp(-i\alpha\gamma_5)\psi_\mu.
\end{array}
\lab{rgravity}
\eeq
{}From the above and Eqs.\eqs{comptrans}{rgaugino} we see that a possible
\R-gauge boson would couple axially to the gravitino, the gauginos, and
the
chiral fermions. It was first noticed by Freedman \cite{freedman} that the
axial gauge vector {\it can} couple to the gauginos and the gravitinos
in an invariant way in {\it local} supersymmetry. The variation of
\eq{rterm} is then cancelled by the variation of the term
\beq
e^{-1}{\cal L}=\frac{i}{\sqrt{2}}
{\overline\psi}_\rho \gamma^{\mu\nu} F_{\mu\nu}^R\gamma^\rho\lam^R,
\eeq
in the action since $\delta\psi_\mu$ contains $g_RV^R_\mu\gamma_5\epsilon$.
Later, Das \etal \cite{das} extended the Fayet-Illiopoulos model \cite{fi}
of global supersymmetry to local supersymmetry. They found that the
abelian
gauge theory was chiral and just that of Freedman \cite{freedman}. These
results \cite{freedman,das} were reproduced in \cite{dewit} including the
gravitational auxiliary fields. Stelle and West \cite{stelle} then derived
the action for the Fayet-Illiopoulos
term in local supersymmetry in the superconformal framework \cite{kugo}
\beq
\int d^4xd^4\theta {E} e^{-\xi g_R V^R},
\lab{fi}
\eeq
where E is the superspace determinant and $\xi$ is the constant of the
FI-term. The expression \eq{fi} is invariant under the \ru gauge
transformations
\barr
V^R&\ra& V^R+\frac{i}{g_R}(\Lambda-{\overline\Lambda}),
\quad \Dbar_{\dot{\alpha}}\Lambda=0,\\
E&\ra& E e^{i\xi(\Lambda-{\overline\Lambda})}.
\earr
The superspace determinant transforms non-trivially.
In Ref.\cite{freedman} it was shown that this implies a \ru charge
$\frac{3}{2}\xi$ for the gauginos and the gravitino. In the previous
section
we had a global \R-charge $+1$ for the gauginos which corresponds to the
choice $\xi=\frac{2}{3}$.

Barbieri \etal \cite{barbstelle} extended this analysis to include matter
fields in a general superpotential. In the superconformal framework
an invariant superpotential is constructed by introducing a compensating
superconformal chiral multiplet $S_0$ which transforms under the \ru
gauge group as
\beq
S_0\ra e^{+i\xi{\Lambda}} S_0, \quad
{\overline S}_0\ra e^{-i\xi{\overline\Lambda}}{\overline S}_0,
\lab{s0}
\eeq
and such that the matter multiplets transform as\footnote{Note that the
chiral fields ${\bar S}_i$ transform with ${\bar\Lambda}$ and not with
$\Lambda$. Thus $g(S_i)g^*(S_i)$ is not \R-invariant.}
\beq
S_i\ra e^{-in_i\xi{\Lambda}} S_i,\quad
{\overline S}^i\ra e^{in_i\xi{\overline\Lambda}}{\overline S}^i,
\lab{rchiral}
\eeq
with \ru  charge $n_i$. Then the action
\beq
\left[ S_0^3 g(S_i)\right]_F,
\eeq
is invariant under the gauge transformations (2.7)-(2.9) provided
\beq
g(S_i)\ra e^{-3i\xi\Lambda}g(S_i).
\lab{rg}
\eeq
The superpotential has a net \ru charge just as in Eq.\eq{rsuper}.
We obtain the same charge $+2$ again for the choice $\xi=\frac{2}{3}$.
Thus it is found that the generalization of the Fayet-Illiopoulos term
to local supersymmetry leads to a gauged \R-symmetry!

In Ref.\cite{ferkugo} Ferrara \etal showed that any \R-invariant gauged
action can be put into the canonical form of local supersymmetry
\cite{cremmer}. The most general Lagrangian with local \R-symmetry (with
not more than 2 derivatives for the component fields) and local
supersymmetry is given by \cite{ferkugo}
\barr
{\cal L}&=&
-\half\left[
{\overline S}_0 e^{-\xi g_R V^R}S_0 \phi
(S_i,({\overline S}^i e^{n_i\xi g_RV^R})e^{2{\tilde g} V})
\right]_D \nonum\\
&&+ \left([g(S_i)S_0^3]_F-[f_{\alpha\beta}(S_i)W^\alpha W^\beta]_F-
[f_R(S_i)W_R^2]_F+ h.c.\right),
\lab{raction}
\earr
where $W_R$ is the field strength of the vector multiplet $V^R$, a
{\it propagating} gauge field, and where the function $\phi$ is invariant
under \eq{rchiral}. $W^\alpha$ is the field strength of other
(non \R-) gauge groups, \eg of $SU(2)_L$. To convert \eq{raction} into
the familiar supergravity form, we first rescale the compensating multiplet
$S_0$
\beq
S_0\ra S_0 g^{-\third},
\lab{s0trans}
\eeq
which reduces the first two terms in \eq{raction} to
\beq
-\half \left[
{\overline S}_0S_0\frac{\phi(S_i,{\overline S}^i e^{n_i\xi g_RV^R}
e^{2{\tilde g}V})}{\left(
g^{*}({\overline S}^i)e^{3\xi g_RV^R}g(S_i)\right)^\third}
\right]_D+[S_0^3]_F.
\lab{rmatact}
\eeq
Using the invariance of the denominator of the first term in
\eq{rmatact} under gauge transformations (including the \R-ones),
the denominator can be rewritten in the form
\beq
\left( g^{*}({\overline S}^i e^{\xi n_ig_RV^R}e^{2{\tilde g}V})g(S_i)
\right)^\third,
\eeq
provided that $g$ satisfies the property \eq{rg}. Here ${\tilde g}$ is the
gauge coupling of the non-R gauge groups. In the minimal formulation of
supergravity the terms in \eq{rmatact} take the form
\beq
-\frac{3}{2} \left[ {\overline S}_0S_0 e^{\third {\sg}}\right]_D +
[S_0^3]_F,
\eeq
which implies that
\beq
e^{\third {\sg}}=\third \frac{\phi(S_i,{\overline S}^i e^{n_i\xi g_RV^R}
e^{2{\tilde g}V})}{ \left(g(S_i)g^{*}({\overline S}^i e^{n_i\xi g_RV^R}
e^{2{\tilde g}V})\right)^\third }\,\, .
\eeq
Then all the results of \cite{cremmer} hold except covariant derivatives
include $V^R$.\footnote{See the addendum in \cite{ferkugo} for the special
case where the superpotential vanishes.} The function ${\cal G}(z_i,
{\bar z}^i)$ can be expressed in the form
\beq
{\sg}(z_i,{\bar z}^i)= 3 \ln (\third \phi(z_i,{\bar z}^i))-\ln|g(z_i)
g^{*}(z^i)|.
\lab{calg}
\eeq
The non-invariance of the term $\ln|g(z_i)g^{*}(z_i)|$ under
\R-transformations implies the appearance of the Fayet-Illiopoulos term in
the potential. This follows because the D-term of the \R-multiplet
\beq
g_R\,{\sg}_{,}^{\,i}\,n_iz_i=g_R(3\frac{\phi^{\,i}_,}{\phi}-
\frac{g_,^{\,i}}{g})n_iz_i.
\eeq
has a constant piece as a consequence of the homogeneity of the
superpotential $g$
\beq
n_iz_ig_,^i=3\xi g.
\lab{fiterm}
\eeq
We will see in Section \ref{sec:susybreaking} that this term is very
important when considering the scalar potential. It leads to a
cosmological constant of order $\kap^4$ which must be cancelled by an
appropriate term. As we will see this fixes the scale of \ru-breaking.

So far we have started with a superpotential $g$, holomorphic in $z_i$
and a K\"ahler potential $\phi$. When constructing an \R-invariant
Lagrangian we explicitly included terms coupling $V^R_\mu$ to all
gauginos (including the \R-gaugino) and the gravitino. We then showed how
$g$ and $\phi$ can be combined to ${\sg}$. For illustration, we reverse
this procedure and start with a $N=1$ locally supersymmetric
action characterised by a function $\sg$ of the form \eq{calg}
and where the superpotential $g(z_i)$ is homogeneous of degree $3\xi$.
To obtain the physical couplings we perform the reverse chiral rotations
\beq
\psi_{\mu L}\ra \left(\frac{g^*}{g}\right)^{1/4} \psi_{\mu L},
\quad
\lam_{L}^\alpha\ra \left(\frac{g^*}{g}\right)^{1/4} \lam_{L}^\alpha,
\quad
\chi_{Li}\ra \left(\frac{g}{g^*}\right)^{1/4} \chi_{Li}.
\lab{chiralrot}
\eeq
Then the coupling of the vector $V^R_\mu$ to the scalars is of the form
\beq
D_\mu z_i=\partial_\mu z_i-g_Rn_iV^R_\mu z_i.
\eeq
The coupling to the spinors $\chi_i$ of the chiral superfields
is given in terms of $\sg$ by
\beq
-{\bar\chi_{Li}}\Dslash z_j \chi_R^k(\sg_{,k}^{\,ij}+\half\sg_{,k}^{\,i}
\sg^{\,j}_{,})
+{\sg}_{,j}^{\,i}{\bar\chi}_{Li}\Dslash\chi_{jR}.
\eeq
For normalized kinetic energies $\sg_{,j}^{\,i}=-\half\delta_i^j$
this gives an effective $V^R_\mu$ coupling to the spinors of the chiral
superfields of \footnote{We neglect here couplings to other gauge fields.}
\beq
D_\mu\chi_{iL}=\partial_\mu\chi_{iL}-ig_R(n_i-\frac{3}{2}\xi)
V_\mu^R\chi_{iL}.
\eeq
The gauginos and gravitinos have ``na\"{\i}ve'' weight zero in the $\sg$
formulation. This can be seen in particular from the rotations
\eq{chiralrot} which apparently change the weights of the fermionic fields.
 Thus the covariant derivative contains no gauge field
$V_\mu^R$. However, the full coupling is given by
\beq
-\half{\bar\lam}\Dslash\lam-\half{\bar\lam}_L\gamma_\mu\lam_R\sg_,^{\,i}
D^\mu z_i =-\frac{1}{2}{\bar\lam}\gamma^\mu \left(D_\mu
\lam_{L}-ig_R\xi\frac{3}{2}V_\mu^R \lam_{L}\right), \lab{gauginor}
\eeq
\barr
e^{-1} \epsilon^{\mu\nu\rho\sigma}\left(-\fourth{\bar\psi}_\mu
\gamma_5\gamma_\nu D_\rho\psi_\sigma +\eighth {\bar\psi}_\mu\gamma_\nu
\psi_\rho\sg_{,j}^{\,i} D_\sigma z_i \right)
&=&\nonum\\
-\frac{1}{4}e^{-1} \eps^{\mu\nu\rho\sigma}
{\bar\psi}_\mu&&\!\!\!\!\!\!\!\!\!\!\!\!\!\!\gamma_5 \gamma_\nu
\left(D_\rho\psi_\sigma-ig_R\xi
\frac{3}{2}V^R_\rho
\psi_{\sigma L} \right).
\lab{rgravitino}
\earr
where we have made use of \eq{fiterm}.
Therefore, as before, the charge of the gauginos and gravitinos is
$\frac{3}{2}\xi$. Note that this applies to all gauginos, including the
\R-gaugino itself. Thus there is a $\lam^R\lam^RV_\mu^R$ coupling
eventhough
\ru is an abelian gauge group. The spinors $\chi_{iL}$ have \R-charge
 $(n_i-\frac{3}{2}\xi)$ and their scalar superpartners $z_i$ have
charge $n_i$.
These numbers coincide with the ones used in the global case for
$\xi=\frac{2}{3}$ so that any term in the superpotential $g$ must satisfy
$\sum n_i=2$ and the superpotential can not contain a constant term
because of \eq{fiterm}. Throughout the rest of the paper we fix the
convention to
\beq
\xi=\frac{2}{3}.
\lab{xiconv}
\eeq

\subsection{Superconformal Approach}
To understand what made gauging the \R-symmetry possible we consider the
embedding in the superconformal approach. The superconformal group has
the generators
\beq
(P_m,\,M_{mn},\,K_m, D),\quad (Q_\alpha,S_\alpha),\quad A.
\eeq
The first set of four generators form  the conformal group of
translations, rotations, conformal boosts and dilatations. The second
set of two are the fermionic generators of supersymmetry and the
``superpartner'' of $K_m$. The last (bosonic) generator $A$ is a continuous
chiral $U(1)$ symmetry. However, there is no corresponding kinetic term
and thus no propagating gauge boson.

Superconformal gravity is based on gauging the superconformal group and
then adding constraints on the field strengths corresponding to
$P_m,\,M_{mn}$, and $S_\alpha$. The constraints are solved for the $M_{mn},
\,K_m,$ and $S_\alpha$ gauge fields and the transformations of the
gauge fields are modified so that the constraints are preserved.
In Eq.\eq{raction} we presented the superconformal action with
an additional $U(1)_R$ gauge group denoted by $W_R$. This action thus
contains {\it two} extra $U(1)$'s beyond those contained in the
$W^\alpha$, namely $U(1)_R$ and the $U(1)_A$ of the superconformal
group.

Multiplets transform under the full superconformal group. The compensating
multiplet $S_0$ in \eq{s0} transforms under both the $U(1)_R$ and
superconformal transformations. Under the chiral $A$ transformations $S_i$
transforms as
\barr
\delta_A z_i &=& \frac{i}{2} n_i z_i\Lambda_A \\
\delta_A\chi_{iL} &=& \frac{i}{2} (n_i-\frac{3}{2}\xi)\chi_{iL}\Lambda_A
\earr
where $z_i$ and $\chi_i$ are the scalar and spinor components of the chiral
multiplets $S_i$. Reducing the superconformal to superpoincar\'{e}
invariance is done by fixing the real and imaginary part of $z_0$,
the spinor $\chi_0$ (the components of $S_0$) and $b_\mu$ (the gauge field
of dilatation). This, however, will also break the $U(1)_R$ invariance.
But a linear combination of $U(1)_R$ and the chiral generator $A$ will
survive; the resulting group we again call $U(1)_R$.
 The transformed $S_0$ in \eq{s0trans} is neutral under the new
$U(1)_R$ gauge group as can be seen from Eqs.\eq{s0} and \eq{rg}. Fixing
the superconformal gauge on the transformed $S_0$  breaks the
superconformal group to the superpoincar\'{e} but leaves the $U(1)_R$
invariant.

\nsect{Conditions for the Cancellation of Anomalies}
\subsection{Family Independent Gauged R-symmetry}
We have seen in the last section that it is only possible to
construct a gauged \R-invariant theory in the framework of
locally supersymmetric theories. To build a realistic model
the new \ru gauge symmetry should be anomaly-free. To be specific
we shall take the $N=1$ locally supersymmetric theory to have the
gauge group
\beq
G_{SM}\times U(1)_R \equiv SU(3)_C\times SU(2)_L\times U(1)_Y\times U(1)_R,
\eeq
which is that of the Standard Model extended by \ru. The matter chiral
multiplets are taken to be the quarks, leptons and a pair of Higgs
doublets with the addition of $G_{SM}$ singlets, $N,$ and $z_m$. These
multiplets are denoted by
\beq
\begin{array}{rlrlrl}
L:& (1,2,-\frac{1}{2},l),\qquad&\Ebar: & (1,1,1,e),\qquad
&Q:&(3,2,\frac{1}{6},q), \\
\Ubar:& ({\bar3},1,-\frac{2}{3},u),\qquad
&\Dbar:& ({\bar3},1,\frac{1}{3},d),\qquad&H:& (1,2,-\frac{1}{2},h),\\
\Hbar:& (1,{\bar2},\frac{1}{2},{\bar h}), \qquad &
N:&  (1,1,0,n),\qquad&
z_m:&  (1,1,0,z_m),
\end{array}
\lab{qmnumbers}
\eeq
where we have indicated in parentheses the $G_{SM}$, and \ru quantum
numbers, respectively. The \ru  quantum numbers are for the chiral
fermions. The bosons will have numbers shifted by one unit, \eg for
the slepton doublet it is $l+1$ ({\it cf.} Eq.\eq{comptrans}).

We shall assume that the superpotential in the observable sector has the
form
\beq
g^{(O)}= h_{E}^{ij}L_i\Ebar_jH + h_{D}^{ij}Q_i\Dbar_jH +
h_{U}^{ij}Q_i\Ubar_j\Hbar+ h_N NH\Hbar,
\lab{superobs}
\eeq
where $h_E,\,h_D,\,h_U$ are the generation mixing Yukawa couplings and
$h_N$ is an additional Yukawa coupling.
So at this stage we assume the theory conserves \R-parity. We have added
the term $NH\Hbar$ instead of $\mu H\Hbar$ as in the MSSM, in order
to incorporate a possible solution to the mu-problem via a vacuum expectation
value $\vev{N}$. The singlets $z_m$ only couple in the hidden sector. The
only requirement that comes from \R-invariance on the form of $g^{(O)}$ is
that it should transform with a global phase under the \R-transformations
as in Eq.\eq{rg}. This implies that
\barr
l+e+h&=&-1, \lab{emass}\\
q+d+h&=&-1,\lab{dmass}\\
q+u+{\bar h}&=& -1,\lab{umass}\\
n+h+{\bar h}&=&-1.\lab{muterm}
\earr
We have employed our convention \eq{xiconv}. The $-1$ corresponds
to $\sum n_i=2$ since we are now considering the fermionic charges,
which are shifted by 1, \ie $\sum (n_i^f+1)=2$ as seen from
Eq.\eq{comptrans}. At this stage we have also assumed that the \R-charges
are family independent,
\eg $l_1=l_2=l_3=l$.

Since the \ru gauge boson is a  propagating gauge boson,
we must consider the relevant anomaly conditions. These severely
constrain the \R-numbers appearing in Eq.\eq{qmnumbers}.
We shall require the $U(1)_R^3$ anomaly, and the mixed
$U(1)_R-U(1)_Y$,  $U(1)_R-SU(2)_L$, and $U(1)_R-SU(3)_C$ anomalies
to vanish. The hypercharge anomalies are satisfied by our choice of
$U(1)_Y$
charges. The equations for the absence of the $U(1)_Y-U(1)_R$ anomalies
give
\barr
C_1\equiv TrY^2R &=&0, \lab{tyyr}\\
TrYR^2&=&0, \\
TrR^3&=&0.
\earr
These can be rewritten in terms of the \R-quantum numbers as
\barr
3[\frac{1}{2}l+e+\frac{1}{6}q+\frac{4}{3}u+\frac{1}{3}d] +
\frac{1}{2}(h+{\bar h}) &=&0, \lab{yyr}\\
3[-l^2+e^2+q^2-2u^2+d^2]-h^2+{\bar h}^2 &=&0, \lab{yrr}\\
3[2l^3+e^3+6q^3+3u^3+3d^3]+ 2h^3+2{\bar h}^3+16
+n^3+\sum z_m^3&=&0. \lab{rrr}
\earr
In the last equation the term $16=13+3$ is due to the
13 gauginos present in our model ($SU(3)_C:8$, $SU(2)_L:3$, $U(1)_Y:1$,
and $U(1)_R:1$) as well as the gravitino. The gravitino contribution
is three times that of a gaugino \cite{gravitino}. As seen in
Eqs.\eqs{gauginor}{rgravitino}, they all have \R-charge 1. The absence of
the mixed $U(1)_R-SU(2)_L$ anomalies implies the condition
\beq
C_2\equiv Tr_{\{SU(2)\}}R=0,
\lab{ttr}
\eeq
where the trace is limited to the non-trivial $SU(2)$ multiplets.
This is evaluated as
\beq
3[\frac{1}{2}l+\frac{3}{2}q]+\frac{1}{2}(h+{\bar h}) +2 = 0.
\lab{sulr}
\eeq
The constant $2$ is due to the $SU(2)$ gauginos. For an arbitrary
group $SU(N)$ the trace over the product of the adjoint representation
generators is just $N$. Similarly the absence of the mixed
$U(1)_R-SU(3)_C$
anomalies implies
\beq
C_3\equiv Tr_{\{SU(3)\}}R=0,
\lab{tthr}
\eeq
where now the trace is limited to non-trivial $SU(3)_C$ multiplets,
\beq
3[q+\frac{1}{2}u+\frac{1}{2}d]+3=0.
\lab{sucr}
\eeq
The cancellation of the mixed gravitational anomaly \cite{gravity} requires
\beq
Tr R=0,
\eeq
where the trace is taken over all states because of the universality of the
gravitational coupling. This implies
\beq
3[2l+e+6q+3u+3d]+2(h+{\bar h})-8+n+\sum z_m=0.
\lab{gravr}
\eeq
The term $-8=13-21$ is due to the 13 gauginos as well as the gravitino. In the
gravitational anomaly the gravitino contribution is $-21$ times the gaugino
contribution \cite{gravitino}.\footnote{We thank D. Castano, D. Freedman, and
C. Manuel for pointing out to us the difference in the anomaly contribution of
a gaugino and the gravitino in Eqs. \eqs{rrr}{gravr}. This was treated
incorrectly in an earlier version of this paper.} To solve the set of ten
equations (\ref{eq:emass}-\ref{eq:muterm}), (\ref{eq:yyr}-\ref{eq:rrr}),
\eqs{sulr}{sucr}, and \eq{gravr} we note that the seven Eqs.
\eq{emass}-\eq{umass}, \eq{yyr}, \eq{yrr}, \eq{sulr}, and \eq{sucr} form a
decoupled system with the seven unknowns $l,e,q,u,d,h,$ and $\bar h$. It is
straight forward to show that these equations are incompatible and do {\it not
} have a solution. This is independent of whether we replace the $NH\Hbar$
term by $\mu H\Hbar$ in the superpotential. Therefore we conclude that when
the \R-numbers of the fields are family independent the \ru extension of the
supersymmetric standard model is anomalous.

There are several ways around this problem of which we shall in turn
discuss three. First, we shall consider whether the anomaly can be
cancelled by the Green Schwarz mechanism \cite{gs}. Second, we shall
consider adding additional fields which transform non-trivially under
$G_{SM}$, and third we shall consider a family-dependent \ru.

\subsection{Green-Schwarz Anomaly Cancellation}
The Green-Schwarz mechanism of anomaly cancellation relies on coupling
the system to a linear multiplet $(B_{\mu\nu},\phi,\chi)$ where
$B_{\mu\nu}$ is an antisymmetric tensor. The field strength of
$B_{\mu\nu}$ is given by
\beq
H=dB,
\eeq
and
\beq
B=B_{\mu\nu}dx^\mu\wedge dx^\nu,
\eeq
is a two-form. In order to cancel the mixed gauge anomalies the action for $
B_{\mu\nu}$ is {\it not} given by $H^2$, which is gauge and Lorentz invariant,
but instead by ${\hat H}^2$. ${\hat H}$ is classically gauge and Lorentz {\it
non}-invariant and is given by
\beq
{\hat H}=H-\omega^{Y.M.}-\omega^L.
\eeq
Here
\barr
\omega^{Y.M.}&=&Tr(AdA+\frac{2}{3}A^3),\\
\omega^L&=&tr(\omega d\omega+\frac{2}{3}\omega^3),\\
A&=&g_aA_\mu^aT^adx^\mu.
\earr
$Tr$ is a trace over the gauge group and $tr$ is a trace over the Lorentz
Clifford algebra. $A$ is a one-form and $T^a$ are the generators of $SU(3)_c$,
$SU(2)_L$, $U(1)_Y$, and $U(1)_R$, while
\beq
\omega=\half\omega_\mu^{ab}\sigma_{ab}dx^\mu,
\eeq
is the spin connection one-form.

The  non-invariant part of the gauge transformations of ${\hat H}^2$ are of
exactly the same form as the mixed gauge anomalies $C_1,C_2,$ and $C_3$ of the
previous subsection. The combined action is gauge invariant, \ie the
transformation of ${\hat H}^2$ cancels the mixed gauge anomalies, provided
\beq
\frac{C_1}{k_1}=\frac{C_2}{k_2}=\frac{C_3}{k_3}.
\lab{gsanomaly}
\eeq
Here the $k_i$ are real constants which take into account the different
normalization of the gauge group generators. In string theories the $k_i$ are
the Ka\v{c}-Moody levels of the gauge algebra. In almost all string models we
have $k_2=k_3$. Most string models have been constructed at level $k=1$ for
non-abelian groups. $k_1$ is not necessarily integer.

For $k_2=k_3$ (not necessarily $=1$) the anomaly cancellation conditions are
compatible if and only if
\beq
C_2=C_1+6.
\eeq
As in reference \cite{ibanez}, we can simplify the equations by assuming that
\beq
\frac{C_2}{C_1}=\frac{3}{5},
\eeq
which corresponds to the choice $\sin^2\theta_w=\frac{3}{8}$ at the
unification scale. In this case
\barr
C_1&=& -15,\\
C_2&=&C_3\,=-9.
\earr
Then the anomaly cancellation equations can all be expressed in terms
of one variable $l'=\frac{30}{7}\cdot l$ beyond the quantum numbers of
the singlet fields $z_m$. The remaining equations are
\barr
-80+\frac{3}{2}l'+\sum z_m&=&0, \lab{gsgrav}\\
-\frac{8004}{9}-24l'+\frac{19}{5} {l'}^2+\frac{3}{8} {l'}^3
+\sum z_m^3 &=&0,
\earr
where we have added the contribution of the linear multiplet $(-1)$
to both equations.

These equations have no rational solution for zero or one singlet.
We have performed a numerical scan for three singlets with charges $m/6,$
and $m$ an integer between $-200$ and $200$ and found no solution.
It is not clear whether the situation would
improve if we allow for different but realistic values for ${C_2}/{C_1}$
as the equations become very complicated. String models have also been
constructed for level-two Ka\v{c}-Moody algebras. We have considered
the cases $k_2=2k_3$ and $k_3=2k_2$. The equations are of similar form
to those above. There are no rational solutions for one or two singlets.

We conclude that it is not possible to cancel the anomaly via the
Green-Schwarz mechanism with a small number of singlets.

\subsection{Non-Singlet Field Extensions}
We want to briefly investigate what possible extensions of the field
content could lead to an anomaly-free family independent \R-symmetry.
First, we consider fields which transform under the electroweak gauge
group. In order to maintain the anomaly cancellation in the Standard Model
we allow for extra generations ($N_g$ is the number of generations) and
pairs of Higgs doublets $N_h$. The seven decoupled anomaly equations lead
to the equation
\beq
N_g=\frac{3N_h}{N_h+3}.
\eeq
This has no positive integer solutions. We thus consider the case of an
extra $SU(3)$-octet chiral superfield $\oc_c$ with
$G_{SM}\times U(1)_R$ quantum numbers $(8,1,0,o_c)$. Octet extensions
have also been considered, for example in \cite{weinberg,hara}. The
anomaly equations for \eq{sucr} and \eq{gravr} change, they are now
\barr
3[q+\frac{1}{2}u+\frac{1}{2}d]+3+3o_c&=&0,\\
3[2l^3+e^3+6q^3+3u^3+3d^3]+ 2h^3+2{\bar h}^3+16
+n^3+\sum z_m^3+8 o_c^3&=&0,\\
3[2l+e+6q+3u+3d]+2(h+{\bar h})-8+n+\sum z_m+8o_c&=&0.
\earr
The seven independent equations are now in eight variables and have a
solution in terms of two variables which we choose to be $l$ and $e$,
\beq
\begin{array}{rclrclrcl}
h&=&-(l+e+1),\quad &{\bar h}&=&l+e-1,\quad &q&=&-\frac{2}{9}-
\frac{1}{3}l,\\
d&=&\frac{2}{9}+\frac{4}{3}l+e,&u&=&\frac{2}{9}-\frac{2}{3} l -e,
 &n&=&1, \qquad \;\;o_c=-1.
\lab{occharg}
\end{array}
\eeq
The remaining equations involving the singlets are then given by
\barr
3(2l+e)-19+\sum_i z_i &=&0,\\
3(2l+e)^3+13+\sum_i z_i^3&=&0.
\earr
For zero or one singlet this has no rational solution.
We performed a numerical scan for singlet charges between -20 and 20
in steps of 1/6. We found no solution for two or three singlets. We
found many (sixty-six) solutions with four singlets. Here we present
three solutions written in terms of the quantum numbers
$(2l+e,z_1,z_2,z_3,z_4)$:
\barr
(1,-\frac{47}{3},-\frac{25}{3},3,13),\\
(\frac{11}{3},-11,\frac{7}{3},\frac{25}{3},\frac{25}{3}),\\
(\frac{1}{3},-20,\frac{20}{3},\frac{47}{3},\frac{47}{3})
\earr
Note in Eq.\eq{occharg} that the fermionic component of the octet chiral
superfield has $R$-charge $-1$ and thus the spin zero component has $R
$-charge $0$.  Therefore when supersymmetry is broken the soft-supersymmetry
breaking terms in the scalar potential involving only the spin zero octet
are unconstrained. Such a potential typically breaks $SU(3)_c$. We do not
consider these solutions any further.\footnote{After we submitted this
paper \cite{cfm} also considered coloured triplets $D=(3,1,-\frac{1}{3})$,
${\bar D}=({\bar 3},1,\third)$ and obtain anomaly-free solutions.}

\subsection{Family Dependent Gauged \ru Symmetry}
The Standard Model has three generations which are only distinguished by
their mass. Clearly this structure requires an explanation. One possibility
is that the difference in the families are explained by a horizontal symmetry
at very high energies. Thus in general we expect at high energies the
electron to have different gauge quantum numbers from the muon or the tau and
similarly for the quarks. Only at low energies are the gauge quantum numbers
in the effective theory family independent. We shall see in
Section~\ref{sec:susybreaking} that \R-symmetries are broken close to the
Planck scale. In accordance with this philosophy we thus expect the
\R-symmetry to be family-dependent as well. In this section we investigate
the conditions for an anomaly-free family-dependent \R-symmetry. We shall
denote the \R-quantum number of the matter fields by $e_i,l_i,q_i,u_i$,
and $d_i$, $i=1,2,3$. Motivated by the successes of the work on symmetric
mass matrices \cite{mass} we shall assume a left-right symmetry for the
matter fields
\beq
Q_R(\chi_{iR}^a)=Q_R(\chi_{iL}^a).\lab{lr}
\eeq
Here $a$ is a flavour index. In particular we have
\beq
e_i=l_i,\quad u_i=d_i=q_i,\quad i=1,2,3.\lab{lrcharg}
\eeq
Also motivated by the structure of the quark and lepton masses we shall
assume that only the fields of the third generation enter the superpotential.
The superpotential for the observable sector is then given by
\beq
g^{(O)}= h_{E}^{33}L_3\Ebar_3H + h_{D}^{33}Q_3\Dbar_3H +
h_{U}^{33}Q_3\Ubar_3\Hbar+ h_N NH\Hbar.\lab{superfamdep}
\eeq
The masses for the first and second generation will be generated
after the breaking of some symmetry, possibly the \R-symmetry.
We shall here not further consider the problem of fermion mass.

The anomaly cancellation equations will keep the same form as
in Eqs.\eq{emass}- \eq{gravr} but with the factor of 3 outside the
quark's and lepton's contributions replaced by $\sum^3_{i=1}$
and Eqs.\eq{emass}-\eq{umass} hold only for the third generation.
Making use of our assumption \eqs{lr}{lrcharg} these equations reduce
respectively to
\barr
\frac{3}{2} (l_1+l_2+l_3) + \frac{11}{6} (q_1+q_2+q_3) + \half (h+{\bar h})
&=&0, \lab{fdyyr}\\
{\bar h}^2-h^2&=&0, \\
3 (l_1^3+l_2^3+l_3^3) + 12 (q_1^3+q_2^3+q_3^3) + 2h^3 +
2{\bar h}^3-8+n^3+\sum z_m^3 &=&0, \lab{fdrrr}\\
\half (l_1+l_2+l_3) + \frac{3}{2} (q_1+q_2+q_3) +\half(h+{\bar h}) +2
&=&0,\lab{fdsut}\\
2(q_1+q_2+q_3)+3&=&0,\lab{fdsuth}\\
3(l_1+l_2+l_3) + 12 (q_1+q_2+q_3) +2 (h+{\bar h})+ 16+n+\sum z_m
&=&0.\lab{fdgrav}
\earr
The requirement that the superpotential has homogeneous weight two gives
\barr
2l_3+h&=&-1,\lab{e3mass}\\
n+h+{\bar h}&=&-1,\\
2q_3+h&=&-1,\lab{u3mass}\\
2q_3+{\bar h}&=&-1. \lab{d3mass}
\earr
Combining  all these equations we get
\beq
\begin{array}{rcll}
h={\bar h}&=&-1,&q_3=l_3=0 ,\\
l_2&=&\frac{5}{2}-l_1,\qquad
&q_2=-(\frac{3}{2}+q_1), \qquad
n=1.
\end{array}
\lab{famdepsolns}
\eeq
The only remaining equations to solve  are \eq{fdgrav} and \eq{fdrrr}
which simplify to
\barr
\frac{45}{2}l_1(l_1-\frac{5}{2})-54q_1(q_1+\frac{3}{2})+\frac{155}{8}+
\sum z^3_m&=&0,\lab{fdcube}\\
\sum z_m&=&\frac{43}{2}.\lab{fdlin}
\earr
We see that at least one singlet must be added. For one extra singlet
we find two independent solutions:
\barr
(q_1,q_2,l_1,l_2)&=&(-\frac{76}{3},\frac{143}{6},-\frac{61}{2},33),\\
(q_1,q_2,l_1,l_2)&=&(-\frac{46}{3},\frac{83}{6},-7,\frac{19}{2}).
\earr
Both solutions have $z=\frac{43}{2}$. In the next section we shall discuss the
breaking of supersymmetry and of the gauged R-symmetry. We will see that this
solution is unsatisfactory in many respects. The charge of the singlet $z$ is
positive which leads to an unacceptable cosmological constant. We also see
that some of the fermionic charges of the observable fields ($q_i,l_i,h,{\bar
h}$) are less than $-1$; the bosonic charges are then negative. The potential
then requires fine-tuning in order to guarantee weak-scale sfermion masses.

For two additional singlets we find many solutions. The two solutions with
the smallest $|q_1|$ values are
\barr
(q_1,q_2,l_1,l_2,z_1,z_2)&=& (-\frac{61}{3},\frac{113}{6},-1,\frac{7}{2},
-6,-\frac{55}{2}),\\
(q_1,q_2,l_1,l_2,z_1,z_2)&=& (\frac{61}{3},-\frac{131}{6},-6,\frac{17}{2},
-7,\frac{57}{2}).
\earr
These solutions have negative singlet charges which makes it possible to
cancel the cosmological constant. However, $q_1$ or $q_2<-1$. We scanned the
three singlet case for appropriate solutions and found one. The fermionic
charges are given by
\beq
\left\{(q_1,q_2,q_3);(l_1,l_2,l_3);(z_1,z_2,z_3)\right\}=
\left\{(-1,-\half,0);(\half,2,0);(-\frac{115}{3},
26,\frac{203}{6})\right\}. \lab{3sing}
\eeq
There are three further physically distinct solutions obtained by the
interchanges $q_1\lra q_2$ and $l_1\lra l_2$. We study this solution in more
detail in the next two sections. However, this solution has very large singlet
charges and has a gauge invariant hidden-sector superpotential with very large
powers of the singlet fields. We thus also studied the four-singlet solutions.

For four singlets we find very many solutions. The solutions with observable
field fermionic charges greater than $-1$ can be classified in two sets of
twelve and ten classes
\barr
q_1\,=\,-1,\quad l_1&=&\frac{n}{6}, \quad n=-6,...,6,\,n\not=0\lab{soludd}\\
q_1\,=\,-\frac{5}{6},\quad l_1&=&\frac{n}{6},\quad n=-6,...,6,\, n\not=-4,0,4
\lab{solgen}
\earr
Physically distinct solutions are again obtained by the interchange $q_1\lra
q_2$ and $l_1\lra l_2$. In \eqs{soludd}{solgen} we have disregarded the
solutions where $l_1=0$. These lead to a further term $L_1H\Ebar_1$ in the
superpotential in contradiction to our assumption of dominant third generation
Higgs Yukawa couplings. As we will see in Section \ref{sec:rp} the solutions
with $q_1=-1$ lead to the simultaneous presence of $L_iQ_j\Dbar_k$ and
$\Ubar_l\Dbar_m\Dbar_n$ in the superpotential. In most cases this leads to an
unacceptable level of proton decay. The exceptions are discussed in
\cite{sherroy}. We thus focus on the solutions \eq{solgen}. In Table~1 we
present the complete charges of one representative of each of the ten classes.

In the next section we shall discuss the breaking of supersymmetry and
R-symmetry for the three singlet solution of Eq.\eq{3sing} and the
four-singlet solutions of Table~1.

\nsect{Supersymmetry and R-symmetry Breaking}
\label{sec:susybreaking}
To have a realistic model both supersymmetry and \R-symmetry must be broken at
low energies. Since we have a locally supersymmetric theory, it is possible to
break supersymmetry spontaneously. The easiest way is to utilize a hidden
sector whose fields are singlets with respect to the Standard Model gauge
group. Depending on whether the \R-symmetry and supersymmetry are to be broken
simultaneously or not, (the bosonic component of) these singlets would have or
not have non-trivial \R-numbers. In the case of a gauged \R-symmetry we have
shown that anomaly free models are not possible for leptons and quarks with
family independent \R-numbers. When we allow for family dependent \R-numbers
for the leptons and quarks while maintaining a left-right symmetry, we obtain
many solutions, including \eqs{3sing}{solgen}.

The \R-number of the superpotential is 2, and a Fayet-Illiopoulos term is
necessarily present in the D-term of the scalar potential. The $g_R$ part of
this is\footnote{Here we have assumed that the kinetic energy is minimal and
of the form \barr y&=&\frac{\kap^2}{2}z_iz^i+...\nonumber\earr}
\beq
g_R^2(\third)^2\left( n_iz^iz_i+\frac{4}{\kap^2}\right)^2,\lab{dterm}
\eeq
and we have a cosmological constant of the order of the Planck scale. In a
realistic model we must avoid giving the squarks and sleptons superheavy
masses \cite{alicl2}, otherwise supersymmetry would be irrelevant at
low-energies. Thus to lowest order the condition
\beq
\vev{n_iz^iz_i}+\frac{4}{\kap^2}=0,
\eeq
must be satisfied. As shown in \cite{cfm}, the first order corrections
will of order $(\kappa m_s^2)^2$, where $m_s$ is the supersymmetry scale
of order $10^2-10^3\gev$.
{}From Eq.\eq{dterm} it should be clear that at least one chiral
superfield must have negative (bosonic) \R-charge. In a realistic model only
the singlets should get a vev at the Planck scale. We thus have the necessary
requirement of a negatively charged singlet (fermionic charge $<-1$). This
lead us in the previous section to reject the one-singlet solution. In the
general minimization of the potential we also expect negatively charged
observable chiral superfields to get a vev. In the previous section we thus
imposed the additional constraint that the observable fields have
semi-positive bosonic charges. This lead us to the three- and four-singlet
solutions.

The bosonic components of the only three-singlet solution are given by
\beq
(z_1,z_2,z_3)=(-\frac{112}{3},27,\frac{209}{6}), \quad q_1=0,\;l_1=
\frac{3}{2}.
\eeq
We have chosen $z_1,z_2,z_3$  such that $z_1<z_2<z_3$. For a realistic model
we must have $\vev{z_1}\approx \oc(\kapi)$. The most general polynomial with
\R-charge 2 for such three singlets is given by
\barr
g'(z_1,z_2,z_3)&=&\frac{1}{\kap^3}\left(a_1\kz{1}^{10}\kz{2}
\kz{3}^{10}+a_2 \kz{1}^{25}\kz{2}^{14}\kz{3}^{16}\right.
\\ &&\left. +a_3\kz{1}^{33}\kz{2}^{7}\kz{3}^{30}+a_4 \kz{1}^{41}
\kz{3}^{44}+ ....\right).
\earr
We have only introduced the Planck scale. We take the arbitrary parameters
$a_k=\oc(1)$.

We can not break supersymmetry via the Polonyi mechanism \cite{polonyi} since
a constant is not \R-invariant. Instead we find the above superpotential
sufficient. When we take at least three non-zero parameters $a_k$ in $g'$ then
it is possible to find solutions for which the total potential $V$ is positive
semi-definite with the value zero at the minimum, and where the D-term is also
zero at the minimum. For this we must of course fine-tune the parameters $a_k
$.

In this case the \R-gauge vector boson mass is of the order of the Planck
mass. The total superpotential is then taken to be of the form
\beq
g=g'(z_1,z_2,z_3)+g^{(O)}(S_i),
\eeq
where $g^{(O)}$ is the observable sector superpotential which only depends on
the Standard Model superfields $S_i$ and is given by \eq{superobs}.

The most general potential  in a locally supersymmetric theory with chiral
multiplets $S_a$ is
\barr
V&=&\frac{1}{\kap^4} e^{{\sg}}\left( {\sg_,^{-1 a}}_b \sg_{,a}
{\sg_,}^b-3\right) +\half {\tilde g}^2 {\cal R}{e} f_{\alpha\beta}^{-1}
\left(\sg_,^a(T^\alpha z)_a\right) \left(\sg_,^b(T^\beta z)_b\right).
\earr
For the three-singlet model we thus obtain the dependence of the \R-symmetry
D-term on $z_1,z_2,z_3$ as
\beq
g_R^2\frac{1}{8}\left(\frac{2}{3}\right)^2\left(
-\frac{112}{3}|z_1|^2+27|z_2|^2+\frac{209}{6}|z_3|^2+\frac{4}{\kap^2}
\right)^2
\eeq
{}From the form of $g'$ it is clear that there is no symmetry in $z_1,z_2,z_3$
and their vevs will be unequal. For the D-term to vanish at the minimum we
must have $|z_2|<\frac{112}{81}$, and $|z_3|<\frac{224}{209}|z_1|$. By
fine-tuning the parameters $a_k$ it might be
possible to arrange for $|z_2|\approx z_3\approx\half|z_1|$ so that $|z_1|
\approx\frac{1}{\sqrt{5}}\kapi$. Then if we start with the natural
Planck scale $\kapi$, the effective value of $g'$ will be
$\frac{m_s}{\kap^2}$, where $m_s=\frac{1}{\kap^2}\left(\frac{1}{5}
\right)^{21}\left(\half\right)^{11}$ is of order $\oc(10^2\gev)$.

To be honest we must stress that studying such potentials is a very difficult
task and needs a careful analysis. We shall assume that $z_1,z_2,z_3\approx\oc
(\kapi)$ with coefficients less than one, so that when these fields are
integrated out one gets $\vev{\kap^2 g'}=m_s$.

By integrating the hidden sector fields $z_1,z_2,z_3$ one obtains the
effective
potential as a function of the light fields $z_i$. The general form of the
effective potential is \cite{alicl}
\barr
V_{eff}&=& |{\hat g}_{,i}|^2 + m_s^2 |z_i|^2 + m_s\left(z_i{\hat g}_{,i}
+(A-3){\hat g}+ h.c. \right) + \frac{1}{8} {\tilde g}^2
({\bar z}^i(T^\alpha z)_i)^2,
\earr
where ${\hat g}$ is related to $g^{(0)}$ through a multiplicative factor
depending on the details of the hidden sector. Similarly for $A$ and $m_s$
which is given by $m_s=\vev{\kap^2g'}$. In the observable sector the singlet
$N$ has \R-number 2, one can show that $N$ can acquire a $\vev{N}={\cal O}(m_s)
$ because it also contributes to the D-term. This can be seen explicitly by
minimizing the effective potential
\barr
V&=& |{\hat g}_{,i}|^2 +m_s^2|z_i|^2+m_s\left(z_i {\hat g}_{,i}
+(A-3){\hat g} +h.c.\right) \nonum \\
&&+\frac{1}{8}g^2\left( H^*\sigma^aH+\Hbar^*\sigma^a\Hbar\right)^2
+\frac{1}{8}{g'}^2\left( H^*H-\Hbar^*\Hbar\right)^2 \nonum \\
&&+\frac{1}{18}g_R^2\left|2|N|^2+\sum_{j=1}^3 q_j(|Q_j|^2+|\Ubar_j|^2
+|\Dbar_j|^2) +l_j(|L_j|^2+|\Ebar_j|^2)\right|^2.
\earr
The above potential has \R-breaking terms present in ${\hat g}$ and
$z^i{\hat g}_{,i}$. Together with the terms in $m_s^2|z_i|^2$ they break
supersymmetry softly. We can use the tree-level effective action plus
the renormalization group equations to find the radiative corrections
and the \R-breaking effects present.

The three singlet solution is problematic with the $\Ubar\Dbar\Dbar$ couplings
as will be clear in the next section. Therefore we must consider the four
singlet solutions which we required to avoid such a problem. The
superpotentials for the ten different classes are given in Table~2.

As before we have to tune the parameters $a_k$ so that the potential is
positive definite and so that $|z_1|,...,|z_4|\approx\oc(\kapi)$ with
coefficients less than one so as to induce a scale such that $\vev{\kap^2 g'}
=m_s=\oc(10^2\gev)$. The effective potential takes the same form as in the
three singlet case, but with different \R-numbers for the squarks and
sleptons.

\medskip

It is possible to add direct gaugino masses because the action contains the
term
\beq
e^{\sg}\sg_{,}^l\sg_l^{-1\,k} f_{\alpha\beta,\,k} {\bar\lam}^\alpha_R
\lambda_R^\beta
\eeq
which for the canonical choice of the kinetic energy becomes
\beq
\fourth e^{\frac{\kap^2}{4}z_iz^i}(g_{,}^i+\frac{\kap^2}{2}{\bar z}^ig)
f_{\alpha\beta,i}{\bar\lam}^\alpha_R
\lambda_R^\beta
\eeq
Thus if we choose
\beq
f_{\alpha\beta}=\delta_{\alpha\beta} f(z_i)
\eeq
where, \eg we can take
\beq
f(z_i)=g'(z_i)
\eeq
which will induce direct gaugino masses of order $\vev{\kap^2g'}
={\cal O}(m_s)$
at the tree level. It is clear that gaugino masses will also be induced
by radiative corrections \cite{barb,fama}.

\nsect{Applications to R-parity Violation}
\label{sec:rp}
When extending the Standard Model to supersymmetry new dimension four Yukawa
couplings are allowed which violate baryon- and lepton-number. When
determining our solutions to the anomaly equations we have explicitly assumed
that the superpotential conserves \R-parity and that all these terms were
forbidden. However, this was merely a working assumption, since we are mainly
interested in an anomaly-free supersymmetric model with a gauged \R-symmetry
and the superpotentials \eq{superobs} (family-independent) or \eq{superfamdep}
(family-dependent) posed the minimal number of constraints. Whether $R_p$ is
conserved or not should only depend on gauge symmetries at the high energy
scale. Therefore, we now investigate which $R_p$ violating terms are allowed
in the anomaly-free models \eqs{3sing}{solgen}.

In order to determine the allowed superpotential terms we must consider the
charge combinations of the leptons and the quarks. We shall denote by $\vec{l}
=(l_1,l_2,l_3)$, and $\vec{q}=(q_1,q_2,q_3)$ the set of family dependent {\it
fermionic} lepton and quark charges. For the three singlet model they are
given in Eq.\eq{3sing}. For the four singlet case we had twenty models with
$\vec{q}=(-1,-\half,0)$ and ten models with ${\vec q}=(-\frac{5}{6},-
\frac{2}{3},0)$. The corresponding leptonic charges are given in Table~3. In
all three- and four-singlet models $h={\bar h}=-1$.

The possible dimension four terms are
\beq
L_iL_j\Ebar_k,\quad L_iQ_j\Dbar_k,
\quad \Ubar_i\Dbar_j\Dbar_k,\quad  {\tilde \mu}L_i\Hbar,
\eeq
where ${\tilde \mu}$ is a dimensionful parameter. The indices $i,j,k$ are
generation indices and we have suppressed the gauge group indices. In the
first term we must have $i\not=j$ due to an anti-symmetry in the $SU(2)_L$
indices. Similarly, in the third term we must have $j\not=k$ due to the
$SU(3)_c$ structure. We have included the last term because the symmetry \ru
distinguishes between the leptonic superfields $L_i$ and the Higgs $H$ and
thus can not be rotated away. In our notation and with the left-right symmetry
the \ru charges of the above terms are given by
\barr
Q_R(L_iL_j\Ebar_k)&=& l_i+l_j+l_k\equiv -1, \\
Q_R(L_iQ_j\Dbar_k)&=& l_i+q_j+q_k\equiv -1, \\
Q_R(\Ubar_i\Dbar_j\Dbar_k)&=& q_i+q_j+q_k \equiv -1, \\
Q_R(L_i\Hbar)&=& l_i+{\bar h}\equiv 0.
\earr
The last equality in each line is the requirement on the fermionic
charges for \ru gauge invariance. The $L_i\Hbar$ term is different
just because we are considering the fermionic charges. The superfield
charges must add to $+2$ for all terms.

For the three singlet solution we obtain the following $G_{SM}\times U(1)_R$
additional dimension-four terms
\barr
LL\Ebar:&& {\rm none}\\
LQ\Dbar:&& \lqd{1}{1}{2},\,\lqd{1}{2}{1};\,\lqd{3}{1}{3},\,\lqd{3}{3}{1},\,
\lqd{3}{2}{2},\\
\Ubar\Dbar\Dbar:&&\udd{3}{1}{3},\,\udd{2}{2}{3},
\earr
$LQ\Dbar$ and $\Ubar\Dbar\Dbar$ terms together can lead to a dangerous level
of proton decay. Recently Carlson, Roy and Sher \cite{sherroy} studied the
proton decay rates from all possible combinations. They found that some of the
above combinations are more weakly bound than expected. But for example the
product of Yukawa couplings for the operators $\udd{2}{2}{3}$ and $LQ\Dbar$ is
restricted to be smaller than $10^{-9}$. We thus exclude the three singlet
solution. Similarly we also exclude the four singlet solutions with $q_1=-1$.
This is the reason why in the previous section we restricted ourselves to the
case $q_1=-\frac{5}{6}$.

For the ten models of Table~1 we find the following sets of gauge invariant
$R$-parity violating dimension-four terms
\barr
I:&&\lle{1}{3}{3},\,\lqd{1}{3}{3} \nonum\\
III:&&\lle{1}{3}{1}\nonum\\
IV:&& \lqd{1}{2}{3},\,\lqd{1}{3}{2},\nonum\\
V:&&\lqd{1}{1}{3},\,\lqd{1}{3}{1},\\
VII:&&\lqd{1}{2}{2},\nonum\\
VIII:&&\lqd{1}{1}{2},\,\lqd{1}{2}{1},\nonum\\
X:&&L_1\Hbar.\nonum
\earr
When determining solutions to the anomaly equations we
had an additional set of solutions under the interchange $l_1\lra l_2$
and $q_1\lra q_2$. We can thus obtain a further set of allowed $R$-parity
violating models
\barr
I':&&\lle{2}{3}{3},\,\lqd{2}{3}{3}\nonum\\
III':&&\lle{2}{3}{2}\nonum\\
IV':&& \lqd{2}{2}{3},\,\lqd{2}{3}{2},\nonum\\
IV'':&& \lqd{2}{1}{3},\,\lqd{2}{3}{1},\nonum\\
IV''':&& \lqd{1}{1}{3},\,\lqd{1}{3}{1},\nonum\\
V':&&\lqd{1}{2}{3},\,\lqd{1}{3}{2},\nonum\\
V'':&&\lqd{2}{1}{3},\,\lqd{2}{3}{1},\\
V''':&&\lqd{2}{2}{3},\,\lqd{2}{3}{2},\nonum\\
VII':&&\lqd{1}{1}{1},\nonum\\
VII'':&&\lqd{2}{1}{1},\nonum\\
VII''':&&\lqd{2}{2}{2},\nonum\\
VIII':&&\lqd{2}{1}{2},\,\lqd{2}{2}{1},\nonum\\
X':&&L_2\Hbar.\nonum
\earr
We find the interesting point that we have models with only $LL\Ebar$ type
couplings, others with only $L_i\Hbar$ or $LQ\Dbar$ couplings. We also
have three sets $II,VI,IX$ where \R-parity is conserved. Thus there is no
logical connection between a conserved \R-symmetry and the status of
\R-parity. They are independent concepts.

The $L_{1,2}\Hbar$ term has a dimensionful coupling $\tilde\mu$
similar to the $\mu$ term of the MSSM. Its natural value in our local
supersymmetric models is $\kap^{-1}$. At low energies, we can rotate
away this term and thus generate $LL\Ebar,LQ\Dbar$ interactions which
are strongly constrained experimentally \cite{barger}. These bounds
translate into ${\tilde\mu}<{\cal O}(m_s)$. In order to avoid a further
hierarchy problem we require the absence of $L_i\Hbar$ terms and therefore
exclude the models ${X},{X}'$.

Interestingly enough, most of the models ($I,I',$ $IV,...,IV'''$, $V,...,
V'''$, $VII,...,VII'''$, $VIII,VIII'$) predict sizeable $L_{1,2}Q_i\Dbar_j$
interactions. The first set leads to resonant squark production at HERA which
has been investigated in detail in \cite{butter}. This should be observable
with an integrated luminosity of about $100\,pb^{-1}$ for squark masses below
$275\gev$. The second set also lead to observable signals at HERA even for
very small couplings as discussed in \cite{mora}.

We point out that only in model $I$ we have additional terms $L_1HN$. These
conserve \R-parity provided $N$ is interpreted as a right-handed neutrino.
$L_1HN$ is a Dirac neutrino mass and requires a very small Yukawa coupling.
We thus exclude model $I$.

It is interesting to note that eventhough for the Higgs Yukawa couplings the
third generation is dominant this is not necessarily the case for the $R_p$
violating interactions.

\nsect{Conclusion}
The purpose of this paper has been to take a first step towards model building
with a gauged \R-symmetry. We have discussed in detail that an \R-symmetry can
only be gauged in local supersymmetry since it does not commute with the
supersymmetry generator. We showed that electroweak extensions of the minimal
superpotential do not lead to an anomaly-free theory, independently of the
number of standard model singlets added. We found anomaly-free
family-independent \R-symmetric models by adding an $SU(3)_c$ octet chiral
superfield. This however typically breaks $SU(3)_c$. We then discussed in
detail the family-dependent anomaly-free \R-symmetry. Making assumptions based
on mass matrix considerations we found solutions with one, two, three and four
additional singlets. We discarded the one- and two-singlet solutions based
on the symmetry breaking pattern. For the three- and four-singlet solutions
we analysed the gauge- and supersymmetry breaking. The $U(1)_R$ symmetry is
necessarily broken near the Planck scale because of the Fayet-Illiopoulos
term. This could naturally lead after symmetry breaking to an expansion
parameter of order the Wolfenstein parameter which is required for a
dynamical generation of the correct mass matrix structure. We generated the
supersymmetry scale of order the weak-scale because of the large powers in the
superpotential. The large powers were determined by the \R-symmetry.

We have allowed for the possibility of a solution to the mu problem via an
additional singlet. But there is no potential for this singlet. We shall
show in \cite{aliherbi2} how a proper solution to the mu-problem can be
obtained.

In the last section we discuss in detail the $R_p$ violating structure of our
models. We find that \R-symmetry and \R-parity are disconnected concepts.
We exclude a large class of our models because they lead to an unacceptable
level of proton decay. The remaining solutions typically predict $LQ\Dbar$
\R-parity violation which could be observed at HERA.

We expect gauged \R-symmetries to be a useful model-building tool in the
future.

{\bf{Acknowledgments}}
\newline
We would like to thank Diego Castano, Dan Freedman and Cristina Manuel
for pointing out an error in the original manuscript in the computation
of the gravitino contribution to the anomaly.

We would like to thank Subir Sarkar for discussions on a possible light
gluino. We would like to thank Lance Dixon, Jean-Pierre Derendinger, Corinne
Heath, Luis Ibanez and Graham Ross for helpful conversations.

\noindent{\bf Note Added:}
After submitting our paper we have received the related work of
Castano, freedman and Manuel \cite{cfm}. We have included a few comments
concerning the connection to their work in text. In particular we have
modified the D-term of the low-energy effective potential.

\begin{table}
\begin{center}
\begin{tabular}{|l|rrrrr|}\hline\hline&&&&&\\[-0.3cm]
&$l_1$ & $z_1$&$z_2 $&$ z_3$ &$ z_4$ \\ \hline\hline &&&&&
\\[-0.3cm]
I&            $ -1$&$-\frac{80}{3}$&$7$&$\frac{53}{3}$&$\frac{47}{2}$\\&&&&&\\
II&$-\frac{5}{6}$&$-\frac{203}{6}$&$21$&$\frac{185}{6}$&$\frac{7}{2}$\\&&&&&\\
III&$-\frac{1}{2}$&$-\frac{131}{2}$&$\frac{95}{6}$&$\frac{391}{6}$&
$6$\\&&&&&\\
IV&$-\frac{1}{3}$&$-24$&$\frac{25}{2}$&$\frac{131}{6}$&$\frac{67}{6}$\\&&&&&\\
V&$-\frac{1}{6}$&$-\frac{133}{6}$&$\frac{27}{2}$&$\frac{115}{6}$&
$11$\\&&&&&\\
VI&$\frac{1}{6}$&$-\frac{263}{6}$&$-\frac{25}{6}$&$36$&
$\frac{67}{2}$\\&&&&&\\
VII&$\frac{1}{3}$&$-\frac{74}{3}$&$\frac{33}{2}$&$\frac{43}{2}$&
$\frac{49}{6}$\\&&&&&\\
VIII&$\frac{1}{2}$&$-\frac{83}{2}$&$\frac{55}{2}$&$37$&
$-\frac{3}{2}$\\&&&&&\\
IX&$\frac{5}{6}$&$-\frac{278}{3}$&$-\frac{17}{2}$&$\frac{187}{6}$&
$\frac{183}{2}$\\&&&&&\\
X&            1&$-\frac{167}{2}$&$24$&$\frac{497}{6}$&$-\frac{11}{6}$\\&&&&&\\
\hline\hline
\end{tabular}
\end{center}
\caption{Fermionic charges of the ten four-singlet solutions with $q_1=-
\frac{5}{6}$. $q_2=-\frac{2}{3}$, $l_2=\frac{5}{2}-l_1$, $q_3=l_3=0$.}
\end{table}

\begin{table}
\begin{center}
\begin{tabular}{|l|cl|}\hline\hline&&\\[-0.3cm]
I& $\kap^3g'=$&$\left(a_1\funfo{2}{2}{2}
+a_2\fun{7}{2}{}{6}\right.$\\
&&$\left. +a_3\fun{8}{2}{5}{4} +a_4\fun{9}{2}{9}{2}\right)$\\
&&\\
II &$\kap^3g'=$&$\left(a_1\fun{4}{4}{}{3}+a_2\funtw{7}{7}{2}
\right.$\\
&&$\left. +a_3 \fun{5}{2}{2}{13}+a_4\funth{6}{7}{10}\right)$\\
&&\\
III&$\kap^3g'=$&$\left(a_1\fun{5}{11}{2}{}+ a_2 \fun{6}{14}{2}{3}
\right.$\\&
&$\left. +a_3 \funtw{5}{3}{18}+a_4\fun{6}{3}{3}{20}\right)$\\
&&\\
IV&$\kap^3g'=$&$\left(a_1\fun{5}{}{4}{}+a_2\funtw{6}{4}{4}
 \right.$\\
&& $\left.+a_3\funfo{7}{7}{3}+a_4\fun{8}{6}{3}{3}\right)$ \\ &&\\
V&$\kap^3g'=$&$\left(a_1\fun{5}{3}{2}{2} + a_2\fun{7}{4}{}{6} \right.$\\
&& $\left.+a_3\funtw{10}{10}{}+a_4\funfo{10}{5}{7} \right)$\\ & &\\
VI&$\kap^3g'=$&$\left(a_1\funth{2}{5}{3} + a_2\funtwth{6}{7} \right.$\\
&& $\left. +a_3\funth{5}{8}{7}+a_4\fun{4}{13}{3}{3} \right)$\\ & &\\
VII & $\kap^3g'=$&$\left(a_1\funth{4}{5}{} + a_2\funfo{9}{2}{8} \right.$\\
&& $\left.+a_3\fun{9}{3}{6}{3}+a_4\fun{9}{4}{4}{6} \right)$\\ && \\
VIII &$\kap^3g'=$&$\left(a_1\funfo{3}{3}{} + a_2\funth{2}{3}{5}\right.$\\
&& $\left.+a_3\fun{6}{2}{5}{4}+a_4 \fun{6}{6}{2}{4}\right)$\\ && \\
IX & $\kap^3g'=$&$\left(a_1\fun{4}{5}{4}{3} + a_2\fun{6}{2}{9}{3} \right.$\\
&& $\left.+a_3\fun{}{17}{4}{}+a_4\fun{3}{14}{9}{} \right)$\\ && \\
X & $\kap^3g'=$&$\left(a_1\funtw{4}{4}{4}+ a_2\fun{10}{3}{9}{3} \right.$\\
&& $\left.+a_3\fun{7}{10}{4}{7}+a_4\funtw{9}{9}{12} \right)$\\ && \\
\hline\hline
\end{tabular}
\end{center}
\caption{\R-invariant superpotentials for the ten different classes of
anomaly-free solutions with $q_1=-\frac{5}{6}$. We have only kept the
lowest four terms.}
\end{table}

\begin{table}
\begin{center}
\begin{tabular}{|l|l|}\hline\hline&\\[-0.3cm]
Model&Lepton Charges\\\hline&\\
I&$\vecl=(-1,\frac{7}{2},0)$\\
II&$\vecl=(-\frac{5}{6},\frac{10}{3},0)$\\
III&$\vecl=(-\half,3,0)$\\
IV&$\vecl=(-\frac{1}{3},\frac{17}{6},0)$\\
V&$\vecl=(-\frac{1}{6},\frac{8}{3},0)$\\
VI&$\vecl=(\frac{1}{6},\frac{7}{3},0)$\\
VII&$\vecl=(\frac{1}{3},\frac{13}{6},0)$\\
VIII&$\vecl=(\half,2,0)$\\
IX&$\vecl=(\frac{5}{6},\frac{5}{3},0)$\\
X& $\vecl=(1,\frac{3}{2},0)$\\
\hline\hline
\end{tabular}
\end{center}
\caption{Leptonic fermionic charges of the ten four-singlet solutions with
$q_1=-\frac{5}{6}$. }
\end{table}

\end{document}